\documentclass[pre]{revtex4}
\usepackage[margin=3cm,a4paper]{geometry}
\usepackage{graphicx}
\usepackage{latexsym}
\usepackage{amsmath}
\usepackage{amssymb}
\usepackage{amsfonts}
\usepackage{amsthm}
\usepackage{epstopdf}
\usepackage{pstricks}
\usepackage{pst-plot}
\usepackage{pstricks-add}
\usepackage{epsf}

\begin{document}
\title{Free surface lump wave dynamics of a saturated  superfluid $^{4}He$  film with nontrivial  boundary condition at the  substrate surface}
\author{ Abhik Mukherjee}

\affiliation{Department of Theoretical Physics and Quantum Technologies, National University of Science and Technology, MISiS, Moscow, Russia}

\begin{abstract} 

 In this article, the free surface wave dynamics of a saturated ($\sim 10^{-6}$ cm) superfluid $^{4}He$  film is considered under the condition that there exists a very weak downward localized superfluid flow 
 into the substrate. For saturated film, the effect of surface tension  plays a decisive role in the surface wave evolution dynamics of the system.  
 The free surface evolution is shown to be governed by forced Kadomtsev Petviashvili-I equation, with the forcing function depending on downward superfluid velocity at the substrate surface. Exact as well as perturbative
 free surface lump wave solutions
 of the (2+1) dimensional nonlinear evolution equation are obtained and the effect of the leakage velocity function on the lump wave solutions are shown.
\end{abstract}
\maketitle

\section{Introduction}	
Upsurge in the research on nonlinear dynamical systems occurred in the last century because of the failure of  linear  theory  in explaining phenomena related to 
large amplitude waves, wave-wave interactions etc \cite{Lakshmanan,Johnson}. But in majority of the cases, the analytical treatment becomes difficult because of
unsolvability of the associated nonlinear evolution equations. Generally, exact solutions are possible to those nonlinear systems which have integrable properties.

Superfluidity is a state of matter when the matter behaves like a fluid with zero viscosity. Study of nonlinear waves in superfluid $^{4}He$ films has started in the last century.
In superfluid $^{4}He$  film, if there exist a small, finite amplitude localized density fluctuation then that can lead to the existence of solitons. These solitons arise from the balance between dispersion 
and nonlinearity in which the nonlinearity emerges from Van der Waals potential of the substrate. Huberman \cite{Huberman} considered monolayer superfluid $^{4}He$ film ($\sim 10^{-7}$)cm to show that small amplitude 
localized perturbations in superfluid density would lead to the occurrence of gap-less solitons. He found this soliton as the solution of Korteweg de-Vries (KdV) equation which is a completely integrable 
system \cite{KdV}. Nakajima et.al \cite{Nakajima1} considered thin $^{4}He$ film and derived KdV equation from Landau-two fluid hydrodynamic approach. In their next paper \cite{Nakajima2} they considered saturated superfluid
$^{4}He$ film ($\sim10^{-6}$ cm) where the surface tension plays a decisive role to the KdV soliton dynamics.  
Biswas and Warke \cite{Biswas1} also derived KdV equation from the phenomenological Hamiltonian given in
\cite{Rutledge} and predicted theoretically that superfluid solitons can exist. Condat and Guyer  \cite{Condat} considered a mixture of $^{4}He$ films and discussed
 the propagation of ``troughlike'' and ``bumplike''
solitons in such films. Johnson \cite{Johnson2} 
on the other hand presented some
linearized theory on  $^{4}He$ films and then discussed  the far-field nonlinear problem. 
He showed that the relevant equation that is valid in the far-field region is the Burgers equation.
In \cite{Gopakumar}, Gopakumar et.al studied  superfluid films with the solitons overtaking collisions 
 and found detailed expression with appropriate correction for
the amplitude dependence of the solitary wave on wave speed. In these studies, the models with (1+1) dimensions have been considered.
Biswas and Warke \cite{Biswas2} considered
(2+1) dimensional model and derived Kadomtsev Petviashvili (KP) equation for surface density fluctuations in superfluid $^{4}He$ film. Later Sreekumar and Nandakumaran \cite{Sreekumar1}
studied two-soliton resonances for the KP equation which was derived as the evolution equation for the superfluid surface density. They also considered
large amplitude density fluctuations in a thin superfluid film and discussed about existence of ``quasi-solitons'' under
collision \cite{Sreekumar2}. In \cite{Sreekumar3}, they showed that the free surface dynamics of a saturated two-dimensional superfluid $^{4}He$  film is 
governed by the KP equation. It is also shown
that  soliton resonance could happen at lowest
order nonlinearity,  if two dimensional effects are considered. In an interesting recent work \cite{PRL}, the first experimental observation of bright soliton has been done in bulk superfluid $^{4}He$.
When the free surface dynamics of $^{4}He$ film
is considered, it is generally assumed that  the superfluid does not flow into the substrate. But there
may be  situations \cite{porous1,porous2} where this bottom boundary condition might change due to a small yet finite downward superflow. The standard trivial bottom boundary condition is used in the derivation of long wave, small amplitude nonlinear integrable equations like KdV, KP equations etc in hydrodynamic systems. In \cite{o1,o2},   such  bottom boundary condition is changed  to include a weak leakage effect in order to  study the solitary wave dynamics.
In this work, we consider a very weak yet finite  downward localized superfluid flow into the substrate. Hence,  nontrivial bottom boundary condition has to be considered in order to study the free surface wave dynamics of the film. We consider a saturated ($\sim10^{-6}$ cm) (2 + 1) dimensional  $^{4}He$ film where the surface tension plays a decisive role 
in the surface wave dynamics. The evolution of the localized lump wave  is studied analytically in presence of such nontrivial bottom boundary condition for the first time in superfluid $^{4}He$ film  as far as our knowledge goes. Nevertheless, our analysis based on superfluid hydrodynamics is admittedly crude because it does not take into account the finer microscopic details of this complicated low temperature system. However, any calculation on such subject  has to start from some simplified analytic study which may be refined later.

This article is organized as follows. In section-II, the derivation of the (2 + 1 ) dimensional nonlinear evolution equation for the free surface of the saturated superfluid $^{4}He$ film with nontrivial bottom boundary condition has been shown. 
Exact as well as perturbative lump wave solutions has been discussed in section-III. 
Finally we conclude our analysis  in section-IV followed by Appendix and bibliography.

\section{Derivation of the nonlinear (2+1) dimensional free surface evolution equation for the saturated superfluid $^{4}He$ film}
\begin{figure}[!h]
\centering

\includegraphics[width =7cm,angle=0]{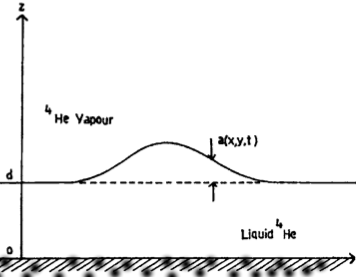}

%
 \caption{\bf A simplified diagram for the propagation of small but finite amplitude  (2 + 1) dimensional localized surface wave in saturated ($\sim10^{-6}$ cm) superfluid $^{4}He$ film with weak downward superflow into the substrate.
 }
\end{figure}
We consider the propagation of small but finite amplitude surface wave on a saturated ($\sim10^{-6}$ cm) superfluid $^{4}He$ film of depth $d$ as  shown in FIG.1. In case of very thin $^{4}He$ film ($\sim10^{-7}$ cm)
the effect of surface tension can be neglected. Nakajima et.al \cite{Nakajima2}
have studied the free surface dynamics of (1 + 1) dimensional  saturated $^{4}He$ film,  the thickness of which is of the order of $~ 10^{-6}$ cm. In such films, 
the surface tension plays a decisive role in the wave dynamics. In this work, we have  considered a saturated (2 + 1) dimensional superfluid $^{4}He$ film with a small yet finite downward localized superflow into the substrate. The downward superflow is localized in the sense that at infinite distance ($x, y\longrightarrow \pm \infty$), it's effect vanishes. 

Any analytical treatment in hydrodynamic system is based on certain assumptions due to the complicated nonlinear characteristics of the system. 
Our calculation relies   on the following considerations:

\begin{enumerate}

\item The acceleration of the superfluid film due to  finite temperature gradient gives a small correction factor \cite{Rutledge}, hence it is neglected. 

\item
Van der Waals force is the nonlinear force acting on the superfluid film \cite{Sreekumar3}. Since, we have considered saturated film ($\sim10^{-7}$ cm), the effect of surface tension is not neglected.

\item The superfluid is assumed to be in-compressible and it's motion to be irrotational.

\item Reductive perturbation technique is used following \cite{Sreekumar3} with the nontrivial bottom boundary condition (\ref{bbc}) to derive the nonlinear dynamical equation. The free surface disturbance $a$ and the equilibrium height $d$ of the film is such that $a/d \ll 1.$

\item We consider that the vertical superfluid velocity at the bottom boundary is very weak so that the interesting surface wave dynamics appear much earlier time, hence the height of the film $d$ does not change significantly. 

\item We consider that the leakage of the superfluid into the substrate is localized in space i.e, it vanishes at space infinities ($x, y \longrightarrow \pm \infty$).

\item For the fast surface wave dynamics, the other physical  parameters like the Van der Waals coefficients are also taken to be constant.

\end{enumerate}


Since we consider irrotational superflow and  in-compressible superfluid  hence from continuity equation we get,
\begin{equation}
 \bigtriangledown^2 \phi(x,y,z,t) = 0,
 \label{Laplace}
\end{equation}
where $\phi(x,y,z,t)$ is the velocity potential \cite{Nakajima2, Sreekumar3}. The system must be supplemented by two variable nonlinear boundary conditions at the free surface and a fixed boundary condition at the bottom \cite{Lakshmanan}.
As discussed before, a small but finite downward flow of superfluid into the substrate is considered. Hence, the bottom boundary condition \cite{Sreekumar3} can be modified as
\begin{equation}
 \frac{\partial \phi}{\partial z}|_{z=0} =  C(x,y,t),
 \label{bbc}
\end{equation}
where $C(x,y,t)$ is the downward superfluid velocity at the bottom surface ($z = 0$) that is dependent on both space(x, y) and time(t) coordinates and vanishes when $x, y \longrightarrow \pm \infty$. To avoid repetitive terminologies,  we will hereafter call this velocity function $C(x,y,t)$ as Damping Function (DF), since it causes damping of the wave energy. The main objective of this work is to find the analytic dependence of the DF on the surface wave profile. 
On the film vapor interface, we have a nonlinear variable boundary condition \cite{Lakshmanan, Sreekumar3} as,
\begin{equation}
 \frac{\partial z_1}{\partial t} + (\frac{\partial \phi}{\partial x})_1 \frac{\partial z_1}{\partial x} + (\frac{\partial \phi}{\partial y})_1 \frac{\partial z_1}{\partial y}- (\frac{\partial \phi}{\partial z})_1 = 0.
 \label{NBC1}
\end{equation}
The index 1 refers to the quantities in the film vapor interface and $z_1 = d + a(x,y,t)$ where $a(x,y,t)$ is the departure of the film surface from the 
equilibrium position.
Another nonlinear boundary condition \cite{Lakshmanan, Sreekumar3}  in the film vapor interface is given as
\begin{equation}
 (\frac{\partial \phi}{\partial t})_1 + 
 (1/2)(\bigtriangledown \phi)^2 -\frac{\sigma}{\rho}(\frac{\partial^2 a}{\partial x^2} +\frac{\partial^2 a}{\partial y^2}) + g_1 a - (1/2)\frac{g_2}{d}a^2  = 0.
\label{NBC2}
 \end{equation}
The last two terms in equation (\ref{NBC2}) comes from the expansion of the Van der Waals force term \cite{Nakajima2, Sreekumar3} with $g_1 = \frac{3 \alpha}{d^4}$ and $g_2 = \frac{12 \alpha}{d^4}$, where $\alpha$  is the Van der Waals constant. $\rho$ and $\sigma$ represent the density and surface tension respectively.
Following \cite{Lakshmanan, Sreekumar3}, we expand $\phi(x,y,z,t)$ in a series as 
\begin{equation}
 \phi(x,y,z,t) = \sum_{n=0}^{\infty} \phi_n(x,y,t) z^n. 
 \label{expan}
\end{equation}
Using (\ref{bbc}) in (\ref{expan}) we get the function $\phi_1(x, y, t)$ as
\begin{equation}
 \phi_1(x,y,t) =  C(x,y,t).
\end{equation}
Following the recursive procedure \cite{Lakshmanan, Sreekumar3}, we can write the velocity potential function $\phi(x,y,z,t)$ as 
\begin{equation}
 \phi(x,y,z,t) = \sum_{n=0}^{\infty}\frac{(-1)^n}{2n!} \phi_{(02n)} z^{2n} +  \sum_{n=0}^{\infty}\frac{(-1)^n}{(2n +1)!} C_{2n} z^{2n +1}, \label{phi}
\end{equation}
where 
\begin{align}
&{}\phi_{(02n)} = \frac{\partial^{2n}\phi_0}{\partial x^{2n}} + \frac{\partial^{2n}\phi_0}{\partial y^{2n}}, \
C_{2n} = \frac{\partial^{2n}C}{\partial x^{2n}} + \frac{\partial^{2n}C}{\partial y^{2n}}.
\end{align}
The second term of (\ref{phi}) comes from the DF (as discussed before) into the substrate.
We study the dynamics of localized surface wave perturbations of long wavelength and small but finite amplitude in the saturated superfluid film thickness. 
Following reductive Perturbation method \cite{Lakshmanan, Sreekumar3}
we expand $\phi_0$, $a$ and $C$ in powers of a small perturbation parameter $\epsilon$ as
\begin{align}
&{} a(x,y,t) = \epsilon a_1(x,y,t) + \epsilon^2 a_2(x,y,t)+ O(\epsilon^3), \\
&{}\phi_0(x,y,t) = \epsilon^{1/2}\phi_0^{(1)}(x,y,t) + \epsilon^{3/2}\phi_0^{(2)}(x,y,t) + O(\epsilon^\frac{5}{2}), \\
&{} C(x,y,t) = \epsilon^{5/2}C^{(1)}(x,y,t) + \epsilon^{7/2}C^{(2)}(x,y,t) + O(\epsilon^\frac{9}{2}),
\end{align}
where $\epsilon$ is a measure of smallness of the perturbation quantities. 
The series of the  term $C(x,y,t)$ is different in powers of small parameter $\epsilon$ because  it's magnitude is assumed to be weak. Using the moving frame transformation \cite{Sreekumar3}
\begin{eqnarray}
 \xi = x + C_3 t, \  t \rightarrow t,
 \end{eqnarray}
where $C_3$ is the frame velocity
and  the asymmetric scaling \cite{Johnson} on the space and time coordinates
\begin{eqnarray}
 \bar{x} = \epsilon^{1/2} \xi, \ \bar{y} = \epsilon y, \ \bar{t} = \epsilon^{3/2}t
\end{eqnarray}
 we get the following equations for the perturbation quantities at the order of  $\epsilon^{3/2}$ and $\epsilon^{5/2}$ respectively  from  equations (\ref{NBC1}) and (\ref{NBC2}). 
 
\begin{align}
 &{} C_3 \frac{\partial a_1}{\partial \bar{x}} + d \frac{\partial^2 \phi_0^{(1)}}{\partial \bar{x}^2} = 0,  
 \end{align}
 \begin{align}
 &{} g_1 \frac{\partial a_1}{\partial \bar{x}} + C_3 \frac{\partial^2 \phi_0^{(1)}}{\partial \bar{x}^2} = 0,  
 \end{align}
 \begin{align}
 &{}C_3 \frac{\partial^2 \phi_0^{(2)}}{\partial \bar{x}^2} - \frac{1}{2} C_3 d^2 \frac{\partial^4 \phi_0^{(1)}}{\partial \bar{x}^4} + \frac{\partial}{\partial \bar{t}} \frac{\partial \phi_0^{(1)}}{\partial \bar{x}} + \frac{1}{2} \frac{\partial}{\partial \bar{x}} (\frac{\partial \phi_0^{(1)}}{\partial \bar{x}})^2 
 -\frac{\sigma}{\rho} \frac{\partial^3 a_1}{\partial \bar{x}^3} + g_1  \frac{\partial a_2}{\partial \bar{x}} - \frac{g_2}{d}a_1  \frac{\partial a_1}{\partial \bar{x}} = 0,  
\end{align}
\begin{align}
&{} C_3\frac{\partial a_2}{\partial \bar{x}} + \frac{\partial a_1}{\partial \bar{t}} + 
 d \frac{\partial^2 \phi_0^{(2)}}{\partial \bar{x}^2} - \frac{1}{6}  d^3 \frac{\partial^4 \phi_0^{(1)}}{\partial \bar{x}^4} + \frac{\partial a_1}{\partial \bar{x}} \frac{\partial \phi_0^{(1)}}{\partial \bar{x}} +a_1  \frac{\partial^2 \phi_{0}^{(1)}}{\partial \bar{x}^2}+  d \frac{\partial^2 \phi_0^{(1)}}{\partial \bar{y}^2} = C^{(1)}.
\end{align}
From  $\epsilon^{3/2}$ order calculations we get 
\begin{equation}
 C_3^2 =g_1 d,
 \label{vel}
\end{equation}
where it is assumed that $a_1,  \phi_0^{(1)} \longrightarrow 0 $ as $x, y \longrightarrow \pm \infty$.
Finally at  $\epsilon^{5/2}$ order, we get the forced Kadomtsev Petviashvili (fKP) equation \cite{Johnson} as the free surface evolution equation having the following form.

\begin{equation}
 A_1 \frac{\partial a_1}{\partial \bar{t}} + A_2  a_1 \frac{\partial a_1}{\partial \bar{x}} + A_3 \frac{\partial^3 a_1}{\partial \bar{x}^3} - 
 A_4 \frac{\partial^2}{\partial \bar{y}^2} \int a_1 d\bar{x} = A_5 C^{(1)},
 \label{fKP}
\end{equation}
where the coefficients are given as
\begin{align}
&{} A_1 = \frac{2}{d}, \ A_2 = (\frac{g_2}{dC_3} - \frac{3C_3}{d^2}), \ A_3 = (\frac{\sigma}{\rho C_3}- \frac{d C_3}{3}), \  A_4 = \frac{g_1}{C_3}, A_5 = \frac{1}{d}.
\end{align}
There are two types of KP equations namely KP-I and KP-II \cite{Johnson} depending on the sign of the coefficients. 
We divide each coefficients of equation (\ref{fKP}) by $A_1$ and redefine new coefficients as 

\begin{align} 
&{} \tilde{A} = \frac{A_2}{A_1}  = \frac{C_3}{2d}, \  \tilde{B} =\frac{A_3}{A_1} = \frac{1}{2 C_3}(\frac{\sigma d}{\rho} - \frac{g_1 d^3}{3}), 
\ 
\tilde{C} = \frac{A_4}{A_1} =  \frac{g_1 d}{2 C_3}, \  \tilde{D} =\frac{A_5}{A_1} = \frac{1}{2}.
\end{align}
The sign of $\tilde{B}$, depends on the size of the thickness $d$, and changes sign at the critical thickness $d_c = \sqrt{\frac{\rho \alpha}{\sigma}}$, 
which in this case is of the order of $10^{-7}$ cm. 
Hence in case of a very thin film
consisting of a few atomic layers \cite{Nakajima1}, the surface tension effect is completely neglected . In this work, we have considered saturated film whose
thickness d is larger than the critical value $d_c$, hence the surface tension plays a vital role in the surface wave dynamics.
This means that coefficients $\tilde{A}$, $\tilde{B}$ and $\tilde{C}$ are all positive.
We can reduce equation (\ref{fKP}) as 
\begin{equation}
  \frac{\partial a_1}{\partial \bar{t}} + \tilde{A}  a_1 \frac{\partial a_1}{\partial \bar{x}} + \tilde{B} \frac{\partial^3 a_1}{\partial \bar{x}^3} - \tilde{C}
  \frac{\partial^2}{\partial \bar{y}^2} \int a_1 d\bar{x} = \frac{1}{2} C^{(1)},
 \label{rfKP}
\end{equation}
where the RHS term ($= \frac{1}{2}C^{(1)}$) comes from the first order perturbation term in DF.
Rescaling variables as $ a_1 = \frac{6 \tilde{B}}{\tilde{A}}U, \  \bar{t} = \frac{T}{\tilde{B}}, \
\bar{x} = X,  \bar{y} = \sqrt{\frac{3\tilde{C}}{\tilde{B}}} Y$ and $C^{(1)} = \frac{12 \tilde{B}^2}{\tilde{A}} f$ we get from (\ref{rfKP})

\begin{equation}
  \frac{\partial U}{\partial T} + 6 U \frac{\partial{U}}{\partial X} +  \frac{\partial^3 U}{\partial X^3} - \int
 3 \frac{\partial^2 U}{\partial Y^2} dX =  - f.
 \label{rrfKP}
\end{equation}
Thus we have got a forced KP equation as the nonlinear dynamical equation for free surface of the film.  We put a negative sign by hand in RHS of (\ref{rrfKP})  because the DF  is directed downward i.e in -z direction. We will refer hereafter, the function $f$ as the scaled Damping  Function (sDF)  for mathematical convenience which vanishes when $X,Y \longrightarrow \pm \infty$. In the following sections, we will explore the dynamical features of $U$ for various functional forms of the sDF  $: f(X, Y, T)$.

\section{Lump wave dynamics}
Since we  considered saturated film ($\sim10^{-6}$ cm) hence  we have got forced KP-I equation because the surface tension effect is dominant which can be seen from (\ref{rrfKP}). It is well known that all completely integrable equations possess soliton solutions \cite{Lakshmanan} which are exponentially localized solutions in certain directions. Whereas lump solutions are  special
kind of rational function solutions which are localized in all directions in the
space. Exact lump solutions are found in many systems such as  KP-I equation \cite{16},
 three-dimensional three-wave resonant interaction system  \cite{17}, 
B-KP equation \cite{18},  Davey–Stewartson-II equation \cite{16}, 
Ishimori-I equation \cite{19} etc. Recently a new completely integrable nonlinear evolution equation possessing lump solution in (2+1) dimension has been developed   \cite{ourpaper,ourpaper2}
which is known as Kundu-Mukherjee-Naskar (KMN) equation \cite{20,21,22,23}.
Since lump wave solutions play an important role in describing localized density/ surface wave fluctuations in various physical systems, we will explore the lump wave dynamics of Eq. (\ref{rrfKP}) for various forcing functions $f$. We know that  N-line soliton states are unstable for KP-I equation whereas  lump solutions are stable \cite{Johnson}. The dynamics of the rational lump wave solutions of Eq.  (\ref{rrfKP}) for different functional forms of $f$ are discussed below.

\subsection{Time dependent sDF  i.e  $f = f(T)$}
Generally, any perturbation to an integrable system spoils it's integrability and exact solvability. However for some specific situations, exact solution is possible when the forcing function   obeys certain  conditions.
Let us  first consider the simple case by considering the downward superfluid velocity into the substrate is time dependent  i.e. $f = f(T)$.
As discussed before, the effect of the function $f(T)$  vanishes at infinite distance
from the region of consideration.
In that case the equation (\ref{rrfKP}) becomes exactly solvable.
We use a shift  $\bar{U} = U + \int f(T) dT$ and translate to a new coordinate $ \tilde{X} = X + a(T) , \tilde{Y} = Y,   \tilde{T} = T$, where $a(T)$ is an arbitrary function of time. If we choose $a(T)$ such that 
\begin{equation}
\frac{da}{dT} = 6 \int f(T) dT \longrightarrow \frac{d^2a}{d T^2} = 6 f(T),
\label{a}
\end{equation}
then equation (\ref{rrfKP}) becomes
\begin{equation}
  \frac{\partial \bar{U}}{\partial \tilde{T}} + 6 \bar{U} \frac{\partial{\bar{U}}}{\partial \tilde{X}} +  \frac{\partial^3 u}{\partial \tilde{X}^3} - 
 3 \frac{\partial^2}{\partial \tilde{Y}^2} \int \bar{U} d\tilde{X} = 0,
 \label{srrfKP}
\end{equation}
which is nothing but  KP-I equation.
The simple one lump solution of (\ref{srrfKP}) is given by
\begin{equation}
 \bar{U} = 4\frac{-\{\tilde{X} + k_1 \tilde{Y} + 3(k_1^2 - m_1^2)\tilde{T} \}^2 + m_1^2(\tilde{Y} + 6 k_1 \tilde{T})^2 + \frac{1}{m_1^2}}{\{\{\tilde{X} + k_1 \tilde{Y} + 3(k_1^2 - m_1^2)\tilde{T} \}^2 + m_1^2(\tilde{Y} + 6 k_1 \tilde{T})^2 + \frac{1}{m_1^2}\}^2},
 \label{lump}
\end{equation}
where $k_1,m_1$ are real parameters. Now moving to the old variables we get,
\begin{equation}
 U = 4\frac{[-\{X + a(T) + k_1 Y + 3(k_1^2 - m_1^2) T \}^2 + m_1^2(Y + 6 k_1 T)^2 + \frac{1}{m_1^2}]}{[\{X + a(T) + k_1 Y + 3(k_1^2 - m_1^2)T \}^2 + m_1^2(Y + 6 k_1 T)^2 + \frac{1}{m_1^2}]^2} - \int f(T) dT,
 \label{olump}
\end{equation}
where $a(T)$ is given by (\ref{a}). Now we will discuss  various situations for time dependent sDF.
\subsubsection[]{\bf $f$ = Constant}
As the simplest case we consider $f$ = constant, which means there is a constant downward superfluid velocity at the bottom surface.
For unforced KP-I equation (when $f = 0$), the 3D plot of the lump solution (\ref{olump}) in the $X-Y$ plane  at a given time is shown in FIG. 2. We can see that the lump wave decreases rapidly as $X, Y \rightarrow \pm \infty$ showing it's localized nature.

 \begin{figure}[!h]
\centering
\includegraphics[width= 7cm, angle=0]{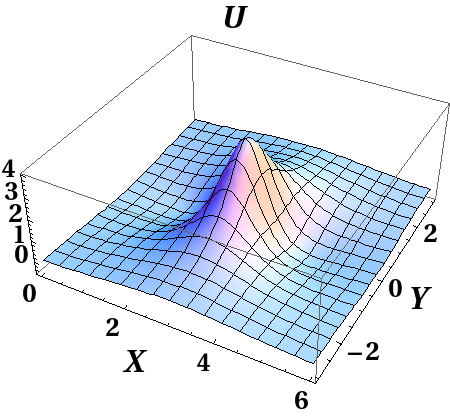}
  \caption{\bf 3D plot of the lump solution (\ref{olump}) in the $X-Y$ plane for unforced KP-I equation ($f = 0$) at  $T = 1$ with $k_1 = 0, m_1 = 1$. We can see that the lump wave decreases rapidly as $X, Y \rightarrow \pm \infty$ showing it's localized nature. 
 }

\end{figure}

\begin{figure}
\includegraphics[width=7cm]{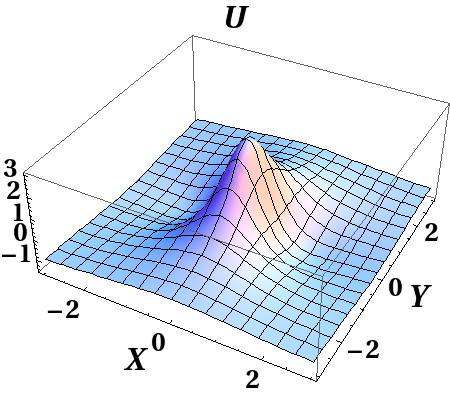}
  \ \ \ \includegraphics[width=7cm]{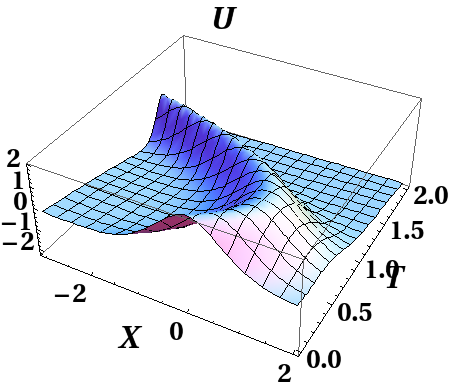}

 (a) 3D plot in  $X-Y$ plane at at T=1
\quad \qquad  (b) 3D plot in $X-T$ plane for Y=1

\vspace{1cm}

\noindent FIG. 3: {\bf 3D plots of the lump solution (\ref{olump}) in the $X-Y$ and $X-T$ plane respectively  for ($f = C_k = 1$) with $k_1 = 0, m_1 = 1$. The dependence of the nonzero constant sDF on the lump solution (\ref{olump}) is clearly seen. }
\end{figure}

For $f = C_k =$ constant,  we  get $a(T) = 3 C_k T^2 + b_1 T + b_2$ from (\ref{a}) where $b_1, b_2$ are arbitrary integration constants which we can neglect here for simplicity. FIG. 3(a) shows the 3D plot of solution (\ref{olump}) in the $X-Y$ plane for $f = C_k =1,$ at $T = 1$. We see that the wave is shifted and it's amplitude is decreased compared with FIG. 2. The effect of this nontrivial sDF can be clearly seen in FIG. 3(b) where the wave is plotted in $X-T$ plane. Due to the time dependent nonlinear function $a(T)$ in the lump wave solution (\ref{olump}) the solution gets curved in $X-T$ plane.

To observe the effect of the constant sDF  ($ C_k$) on the free surface lump wave solution (\ref{olump}), we have to concentrate on the  solution (\ref{olump}). For the choice of parameters $k_1=0, m_1 = 1$ (which we have chosen in the previous plots) we can write down the solution (\ref{olump}) as
\begin{equation}
 U = \frac{4[-\{X + (3 C_k T  - 3 ) T \}^2 + Y^2 + 1]}{[\{X + (3 C_k T - 3) T \}^2 + Y ^2 + 1]^2} - T C_k
 \label{olump1},
\end{equation}
where $a(T) = 3 C_k T^2$ for constant $f = C_k$, as we have discussed before.  At infinite distance from the region of observation, we can neglect the effect of this constant sDF.

\begin{enumerate}
 \item 
 We can see an interesting feature of the lump wave dynamics from Eq. (\ref{olump1}). We see that the velocity of the wave is time dependent i.e, $V(T) = 3(C_k T - 1)$. When $T$ increases from zero,  the lump wave propagates along positive direction of $X$ axis.
 
 \item
 For $T_c = \frac{1}{C_k}$ (for the given choice of parameters), the velocity becomes zero and the wave stops instantaneously.
 
 \item
 For further increase of time, the wave moves towards opposite direction i.e, towards negative $X$ axis and continues its motion.
 
 \item
 Now if we concentrate on the wave fluctuations at the origin i.e at $X = Y = 0$, we can get the plots of FIG. 4 for different constant values of $C_k.$ We can see that, when $T$ increases the amplitude decreases at origin ($X=Y=0$) because the wave moves away from origin towards positive $X$ axis. After $T = T_c$ as we have discussed, the wave reverses it's motion and starts to come towards origin, then the amplitude at origin again increases (described by the hump structures in plots $4(b)-4(d)$). After crossing origin the wave moves away towards negative $X$ axis, hence the amplitude at origin again decreases as shown in the plots. Depending on the values of $C_k,$ the reflection of wave occurs at different time $T_c$ which are shown in plots $4(b)-4(d)$.
 
 \item 
 A decaying background  term ($-T C_k$) appears in (\ref{olump1}) which comes from the second term of Eq. (\ref{olump}). It is not interesting because it does not propagate with time. It can be explained as, the wave height $U$ continuously decreases due to a constant leakage ( $f = C_k$) of superfluid into the substrate. Hence due to this secular background term the equilibrium wave height which is $\gg$ U, decreases significantly at large time.

\end{enumerate}

\begin{figure}

\includegraphics[width=7cm]{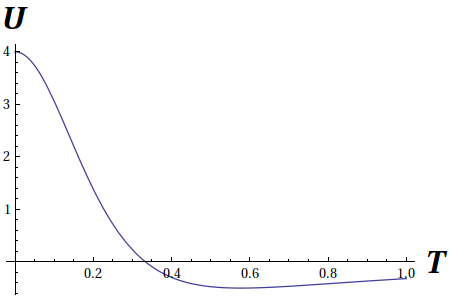}
 \ \ \ \includegraphics[width=7cm]{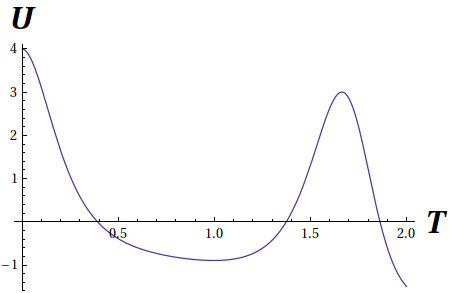}

\vspace{1cm}

 (a) $f = C_k = 0$ 
\quad \qquad \quad \qquad\quad \qquad(b)   $f = C_k = 0.6$ 

\vspace{1cm}

\includegraphics[width=7cm]{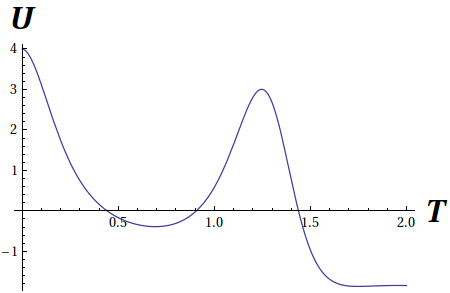}
 \ \ \ \includegraphics[width=7cm]{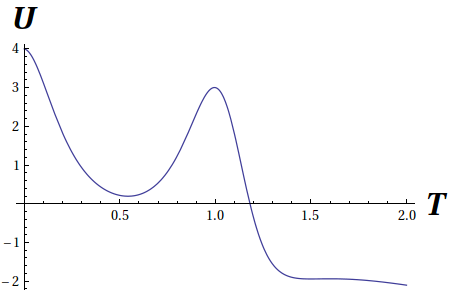}

\vspace{1cm}
(c)  $f = C_k = 0.8$ 
\quad \qquad \quad \qquad\quad \qquad(d)   $f = C_k = 1$ 

\vspace{1cm}

\noindent FIG.4 : {We plot the lump wave solution (\ref{olump1}) at the origin ( $X=Y=0$)  with time $T$ for various values of the constant $C_k$ with $k_1 = 0, m_1=1$. It is clear from the figure that the wave moving along positive $X$ axis  gets reversed at time $T=T_c = 1/C_k$ and continues to move along negative $X$ axis as discussed in the main text. Hence the wave amplitude at the origin decreases and then again increases which is shown by the hump like structures in the plots. }
\end{figure}

\subsubsection{\bf $f$ = $ \sin{(\Omega T)}$ where $\Omega$ is constant parameter}
\begin{figure}[!h]
\centering
\includegraphics[width= 7cm, angle=0]{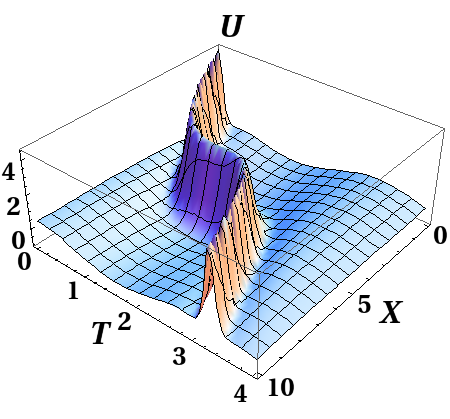}

\vspace{1cm}
 \noindent FIG. 5: {3D plot of the solution (\ref{olump}) in $X-T$ plane at $Y = 1$ for $a(T) = -\frac{6}{\Omega^2} \sin{\Omega T}$ with $k_1 = 0, m_1=1$. It is seen from the plot that the wave changes its direction of motion continuously along $X$ axis since it's velocity depends on the sinusoidal function $a(T)$ . }
\end{figure}
 \begin{figure}
\includegraphics[width=7cm]{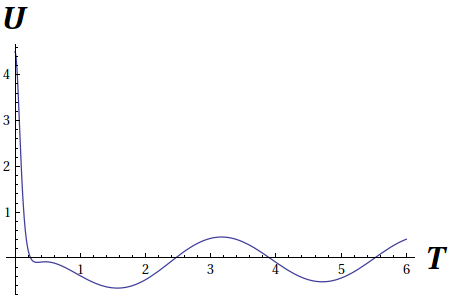}
  \ \ \ \includegraphics[width=7cm]{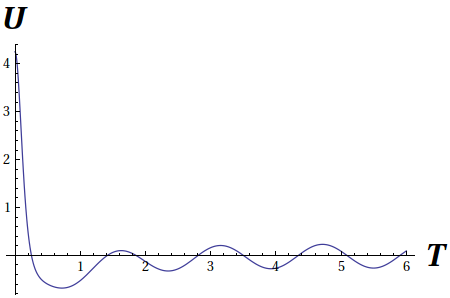}

 (a) Plot of $U(X=Y=0)$ vs T for $f = \sin{2T}$
\quad \qquad  (b) Plot of $U(X=Y=0)$ vs T for $f = \sin{4T}$

\vspace{1cm}

\noindent FIG. 6: {We plot the lump wave solution (\ref{olump}) at the origin ( $X=Y=0$)  with time $T$ for $f = \sin{\Omega T}$ ($\Omega = 2,4$)  with $k_1 = 0, m_1=1$. It is clear from the figure that the wave reverses it's direction of motion continuously along $X$ axis.   Hence the wave amplitude at the origin also oscillates  which is shown in the plots.}
\end{figure}
Now as an extension to the previous study, we consider a time dependent sDF. 
Depending on the real physical condition, the functional form of $f$ may be different. As an example we consider a sinusoidal term as
 $f =  \sin{(\Omega T)}$ where $\Omega$ is a constant parameter. Then the function $a(T)$ can be calculated from (\ref{a}) as $a(T) = -\frac{6}{\Omega^2} \sin{\Omega T}$ where arbitrary constants have been neglected like before. FIG. 5 shows the 3D plot of the lump wave solution (\ref{olump}) in the $X-T$ plane for $Y = 1$. Due to the sinusoidal form of $a(T)$, the velocity of the wave changes according to this oscillating function. Hence the wave continuously changes it's direction of motion which can be seen from FIG. (5). Similar to FIG. 4, we plot the wave dynamics at $X=Y=0$ for two values of $\Omega.$

We can choose different functional forms of $f(T)$ depending on the porous structure of the substrate. The sDF should be chosen accordingly to explain the real wave dynamics. Obviously such effect of sDF on the surface lump wave profile $U$ vanishes at infinite distance i.e, when $X,Y \longrightarrow \pm \infty$.

\subsection{Space - time dependent sDF  i.e  $f = f(X,Y,T)$}
Finally, we discuss the most general case that is
when the downward velocity function $f$ depends on both space-time coordinates ($X, Y, T$). In that case, the exact solution of (\ref{rrfKP}) becomes difficult. Although, it can be solved for a situation when the forcing function satisfies a definite nonholonomic constraint.
Soliton equations with self-consistent sources are important problems in many branches of physics and applied mathematics. The special lump wave solution of (\ref{rrfKP}) can be evaluated following \cite{Yong} where the external forcing function satisfies a constraint. But in real experimental conditions, such solutions are not very relevant because the wave and the sDF does not usually satisfy such kind of constraint. Nevertheless, the solution is interesting in terms of mathematical point of view, hence the derivation of the solution is shown in Appendix. 

For obtaining a general, physically relevant lump wave solution of (\ref{rrfKP}), we use  perturbation method which is discussed below.  
For a  general choice  of  $F ( = \frac{\partial f}{\partial X})$, Eq. (\ref{rrfKP}) can be solved via perturbation technique if the downward velocity function $F (X,Y,T)$ is assumed to be fast compared to the evolution scale of the unforced equation. This suggests the introduction of two time scales in the mathematical procedure.  The effect of rapid forcing on some  evolution equations has been investigated in \cite{1,2,3}. In \cite{3}, two-dimensional perturbed KP equation has been considered and solved for general initial conditions and forcing functions. We follow the method from \cite{3}  to derive the full perturbative solution of our system (\ref{rrfKP}).

\subsubsection{\bf Perturbation method for rapidly varying superfluid velocity function, $F$ }
We assume that the  function, $F$ (which is the space derivative  of sDF into the substrate i.e, $f_X$)  is a rapidly varying function i.e, $F$ = F(X,Y,T/$\epsilon_1$), where $ 0 < \epsilon_1 << 1 $. Following \cite{2}, we introduce a fast time scale, $\tau = T/\epsilon_1$ in the calculation where $\epsilon_1$ is a small parameter.
In that case, the time derivatives transform according to the rule  $\frac{\partial}{\partial T}$ $\rightarrow (1/\epsilon_1)$ $\frac{\partial}{\partial \tau}$ + $ \frac{\partial}{\partial T}$. 
Applying these transformations to (\ref{rrfKP}) we get
\begin{equation}
  ( U_T + 6 U U_X + U_{XXX})_X - 3 U_{YY} + \frac{1}{\epsilon_1}U_{X \tau} = -F (X, Y, \tau),
\label{prrfKP}
\end{equation}
where subscripts denote partial derivatives. We expand the wave profile, $U$ in a series of small parameter $\epsilon_1$ as,
\begin{equation}
 U(X,Y,T,\tau) = \sum_{n=0}^{\infty} U_n(X,Y,T,\tau) \epsilon_{1}^n. 
 \label{expan2}
\end{equation}
Transforming (\ref{prrfKP}) accordingly and equating coefficients of different powers of $\epsilon_1$ to zero we get the following evolution equations for perturbation quantities.

\vspace{0.5cm}

$O(1/\epsilon_1)$ :
\begin{equation}
 U_{0X\tau} = 0.
 \label{-1}
\end{equation}

$O(1)$ :
\begin{equation}
 U_{0XT} + 6(U_0 U_{0X})_X + U_{0XXXX} - 3U_{0YY} + U_{1X\tau} = - F.
 \label{0}
\end{equation}

$O(\epsilon_1)$ :
\begin{equation}
 U_{1XT} + 6(U_0 U_{1X})_X + 6(U_1 U_{0X})_X+ U_{1XXXX} - 3U_{1YY} + U_{2 X \tau} = 0.
 \label{1}
\end{equation}
Our aim would be to solve these perturbation quantities at each order to evaluate the full perturbative solution.
Integrating Eq. (\ref{-1}) twice we get,
\begin{equation}
 U_0(X,Y,T,\tau) = V_0(X,Y,T) + W(Y,T,\tau),
\end{equation}
where $V_0, W$ are two new functions to be determined later.  For lump solutions, the wave vanishes as $X \rightarrow \pm \infty$ with fixed Y making $W(Y,T,\tau) = 0$. Now we can write
\begin{equation}
 U_0(X,Y,T,\tau) = V_0(X,Y,T),
 \label{U0}
\end{equation}
where $V_0(X,Y,T)$ will be determined in next order. We can see that the zeroth order wave profile $U_0,$ does not depend on the fast time scale $\tau$. This is evident because when $\epsilon_1 = 0,$ there is no  forcing function hence we get the lump solution of unperturbed KP equation.
From (\ref{0}) we can separate out two equations as $\tau$ independent and $\tau$ dependent parts respectively using (\ref{U0}) as,
\begin{equation}
 V_{0XT} + 6(V_0 V_{0X})_X + V_{0XXXX} - 3V_{0YY} = 0,  \label{V0}
 \end{equation}
 \begin{equation}
 U_{1X\tau} = - F(X,Y,\tau).
 \label{sol0}
 \end{equation}
 Thus we see that $V_0$ satisfies unforced KP-I equation.
Now, solving (\ref{sol0}) we get,
\begin{equation}
 U_1(X,Y,T,\tau) = V_1(X,Y,T) + G_1(X,Y,T,\tau), \ \ G_1(X,Y,T,\tau) = - \int\int F(X,Y,\tau)dX d\tau, \label{G1}
\end{equation}
where $V_1(X,Y,T)$ will be determined in next order.
Hence the forcing function $F$ comes into the solution at $O(\epsilon_1)$.
From (\ref{1}) we again separate out two equations as $\tau$ independent and $\tau$ dependent parts respectively using previous results as,
\begin{eqnarray}
 V_{1XT} + 6(V_0 V_{1X})_X + 6(V_1 V_{0X})_X + V_{1XXXX} - 3V_{1YY} = 0, \label{V1} \\
  G_{1XT} + 6(V_0 G_{1X})_X + 6(G_1 V_{0X})_X + G_{1XXXX} - 3G_{1YY} + U_{2X\tau} = 0. \label{G1XT}
  \label{sol1}
 \end{eqnarray}
Differentiating (\ref{V0}) w.r.to X, we can easily see that (\ref{V1}) is identical if we take $V_1 = V_{0X}$. 
Integrating (\ref{G1XT}) we get,
\begin{equation}
 U_2(X,Y,T,\tau) = V_2(X,Y,T) + G_2(X,Y,T,\tau),
 \end{equation}
 where 
 \begin{equation}
 G_2(X,Y,T,\tau) = -\int \{ G_{1T} + 6(V_0 G_{1X}) + 6(G_1 V_{0X}) + G_{1XXX} - 3\int G_{1YY} dX \} d\tau. \label{G2}
\end{equation}
Following the same procedure we can get $V_2(X,Y,T)$ in the next order as $V_2 =\frac{1}{2} V_{0XX}$. Thus we can  finally summarize the solutions at different orders in the following manner:
\begin{align}
 &{} U_0(X,Y,T,\tau) = V_0(X,Y,T), \\
 \nonumber \\
 \nonumber \\
&{} U_1(X,Y,T,\tau) = V_1(X,Y,T) + G_1(X,Y,T,\tau), \nonumber \\
&{} G_1(X,Y,T,\tau) = - \int\int F(X,Y,\tau) dX d\tau,   \ \ V_1(X,Y,T) = V_{0X}, \\
  \nonumber \\
  \nonumber \\
&{}  U_2(X,Y,T,\tau) = V_2(X,Y,T) + G_2(X,Y,T,\tau), \nonumber \\
&{} G_2(X,Y,T,\tau) = -\int \{ G_{1T} + 6(V_0 G_{1X}) + 6(G_1 V_{0X}) + G_{1XXX} - 3\int G_{1YY} dX \} d\tau, \nonumber \\
&{} V_2(X,Y,T) = \frac{1}{2} V_{0XX}.
 \end{align}
 Hence the solutions at different orders depend on the unperturbed KP-I solution  and on the  function $F$.
  Till now we have not discussed the role of the choice of initial conditions to the solutions. We will simplify the expression of full perturbative solution for a given initial condition and forcing function in the next subsection.

 \subsubsection{\bf Role of initial conditions}
Let us  consider that  Eq. (\ref{prrfKP}) is to be solved with an initial condition (IC) : $U(X,Y,0,0) = g(X,Y)$.
In \cite{3}, authors have discussed in detail how the initial conditions and the system functions must be self consistently related to each other. In that paper,
they have deduced that when the initial data enter at leading order only, inconsistencies can arise in the higher-order problems. On the other hand 
when the initial data influences all terms in the perturbation expansion, such inconsistencies can be
avoided by an appropriate choice of initial conditions at each order. For a more general treatment, we have considered the second case  where the initial data influences all terms in the perturbation expansion such that

\begin{equation}
 U_j(X,Y,0,0) = g_j(X,Y), j \geq 0,
\end{equation}
where $g_1(X,Y), g_2(X,Y)$ are ICs at first and second order respectively.
Let us consider that the forcing function is of the form 
\begin{equation}
 F(X,Y,\tau) = R(X,Y) S(\tau).
 \label{F}
\end{equation}
In this work, as an example we will consider a given form of $s(\tau)$ to calculate the full perturbative solution.
Hence, from (\ref{G1}) we get
\begin{equation}
 \int R(X,Y)dX = \frac{V_{0X}(X,Y,0)}{\int S(\tau)d\tau|_{\tau=0}}.
\end{equation}
If we chose $S(\tau) = \sin{(\Gamma \tau)}$ where $\Gamma$ is constant, then we can get ultimately,
\begin{equation}
 R(X,Y) = - \Gamma \  V_{0XX}(X,Y,0).
\end{equation}
Hence we can write the  expression of the forcing function from (\ref{F}) as 
\begin{equation}
F(X,Y,\tau) = - \Gamma \ V_{0XX}(X,Y,0)\sin{\Gamma \tau}, \ f(X,Y,\tau) = - \Gamma \ V_{0X}(X,Y,0)\sin{\Gamma \tau}.
\label{F1}
\end{equation}
Thus we can evaluate the functions $G_1, G_2$ appearing in (\ref{G1}), (\ref{G2}) respectively for the specific form of $F$ derived in (\ref{F1}) as
\begin{equation}
 G_1(X,Y,T,\tau) = V_{0X}\cos{\Gamma \tau}, \ G_2(X,Y,T,\tau) = 0. \label{G1G2} 
\end{equation}
Thus we can identify  that the ICs at first and second orders are, $g_1(X,Y) = 2 V_{0X},\  g_2(X,Y) =\frac{V_{0XX}}{2}$ respectively i.e, self consistently related to the solution of unforced KP-I equation as discussed in \cite{3}.

\subsubsection{\bf Full perturbative solution}
Finally we can write the full perturbative solution up to $O(\epsilon_1^2)$ as
\begin{equation}
 U(X,Y,T,\tau) = [ V_0(X,Y,T) + \epsilon_1 \{ V_{0X}(1 + \cos{\Gamma \tau})\} + \epsilon_1^2 \{\frac{1}{2}V_{0XX}(X,Y,T)\} + O(\epsilon_1^3) ],
 \label{solution}
 \end{equation}
where $V_0(X,Y,T)$ is the solution of unforced KP-I equation. Similarly the downward leakage velocity function under consideration i.e, $F$ or $f$ as defined in (\ref{rrfKP})  is given in Eq. (\ref{F1}). Similar to our previous analysis we plot the perturbative solution (\ref{solution}) at the origin ( $X=Y=0$)  with time $T$ in FIG. 7 for $f = 0$ (unperturbed case) and $f =- \Gamma \ V_{0X}(X,Y,0)\sin{\Gamma \tau} $ ($\Gamma = 2,4$)   as given in (\ref{F1}).
\begin{figure}

\includegraphics[width=7cm]{f0.png}
 
\vspace{0.5cm}

 (a) $f = 0$

\includegraphics[width=7cm]{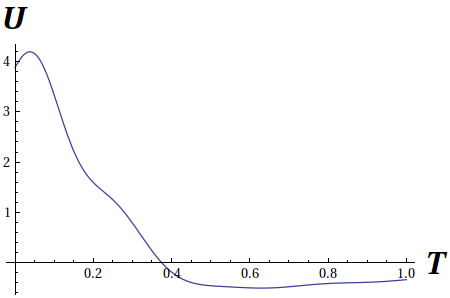}

\vspace{0.5cm}

(b) $f(X,Y,\tau) = - 2 \ V_{0X}(X,Y,0)\sin{2 \tau}$

\includegraphics[width=7cm]{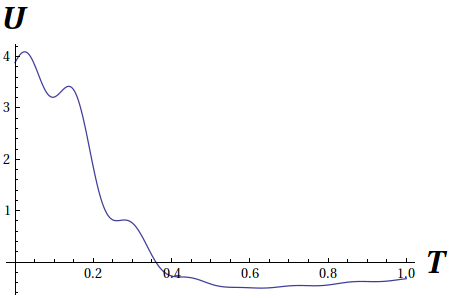}

\vspace{0.5cm}

 (c) $f(X,Y,\tau) = - 4 \ V_{0X}(X,Y,0)\sin{4 \tau}$

\vspace{1cm}

\noindent FIG.7: {We plot the full perturbative solution of the lump wave (\ref{solution}) at the origin ( $X=Y=0$)  with time $T$ for $f = 0$ (unperturbed case) and $f =- \Gamma \ V_{0X}(X,Y,0)\sin{\Gamma \tau} $ ($\Gamma = 2,4$)  with $k_1 = 0, m_1=1$ as discussed in (\ref{F1}). Due to the sDF with the form given in (\ref{F1}) the perturbative wave profile at origin fluctuates as seen from the figures. This fluctuations actually describe the continuous change of direction of motion of the wave along $X$ axis due to sDF.}
\end{figure}

\section{Discussions and conclusive remarks}

Superfluid films are generally described by a two fluid model containing two interpenetrating  fluids : a normal fluid component and a superfluid component. Each fluid components are governed by different dynamical equations. The superfluid component flows without viscosity.
In previous scientific literatures \cite{Nakajima1,Nakajima2}, it is proved that  surface waves can exist in superfluid thin $^{4}He$  films, and those waves are
called ``third sound''.   
In that case, the superfluid component oscillates
parallel to the substrate while the normal fluid component is
held stationary by viscosity. Measurement of ``third sound'' is one of the important technique in this field to determine various properties of the film. It is shown in \cite{Nakajima2}, that the measurement of ``third sound'' becomes considerably easier for saturated film compared to thin film due to smaller wave velocity. In this work, by introducing an external leakage function $f$ which is referred as sDF in the text, we have achieved to change the velocity of the lump surface wave. Hence by a suitable choice of $f$ in this model, we can ease the ``third sound ''
measurement by decreasing the wave speed. In previous scientific literatures, the surface wave dynamics of saturated superfluid $^{4}He$ film has been considered in very few cases  \cite{Nakajima2,Sreekumar3}.
In \cite{Nakajima2}, KdV equation was derived as surface wave evolution
equation where thermomechanical force was neglected.
The resulting solitary waves derived from KdV equation were ``cold'', in contrast to the solitary wave of
very thin film. Whereas in \cite{Sreekumar3}, KP equation was derived in case of saturated superfluid film with the inclusion of weak transverse effects.
But in all such cases trivial bottom boundary condition has been considered such that there is no downward superflow into the substrate. But in case of porous substrates \cite{porous1,porous2}
such conditions must change. 
In shallow water long wave systems in (1+1) dimensions, such nontrivial bottom boundary condition has been  considered in \cite{o1,o2}. Phase and amplitude modifications of 
solitary waves of perturbed KdV equation has been obtained analytically. But in present problem, two dimensional effects are present and surface tension plays an effective role in the dynamics
which is neglected in long water wave system. So, the effects of nontrivial bottom boundary conditions on the dynamics of 
lump solutions of superfluid $^{4}He$ film has not been evaluated analytically till now as far as our knowledge goes.
Hence, those derived lump solutions in this article may be useful in relevant practical  situations. Although we admit that for a more accurate  modeling of such complicated system, the microscopic picture of the fluid - substrate interface like the pore size, pore pressure etc should be taken into account.  In case of a weak yet finite vertical superflow, the Van der Waals coefficient, depth of the film etc should be time dependent. Also the effect of temperature gradient  may affect the wave dynamics. As an initial attempt to this new problem we have used a simplistic condition and used the technique of superfluid hydrodynamics. Such analytic study may be refined later in order to consider more realistic situation.

 In conclusion we can say that, we have considered saturated superfluid $^{4}He$ film, with nontrivial bottom boundary conditions. In such situations, there is a very weak yet finite superfluid velocity
 into the substrate. For example, Nuclepore is a polycarbonate sheet which is around 10 micron thick, can be used as  porous substrate For $^{4}He$  film. In such cases, due to a very weak superflow into the
 substrate the bottom boundary condition will change, which will affect the dynamics of the surface waves.
 In case of saturated films, the effect of surface tension plays a decisive role in the surface wave propagation. We have shown that, in presence of very weak bottom boundary conditions, 
 the dynamics of the free surface is  governed by forced Kadomtsev Petviashvili-I equation, with the forcing function depending on downward superfluid velocity in the substrate. 
 Since, for KP-I equation, line solitons are unstable where as rational lump solutions are stable, so we have tried to concentrate on the dynamics of the simple
 one lump solution of obtained evolution equation. If the downward superfluid velocity depends only on time, then the system becomes exactly solvable. We have shown
 how such weak superfluid velocity affects the structure of lump solutions. If the leakage velocity depends on both space and time co ordinates, then exact solution
 becomes difficult. It is solved by pertubative technique, when there exist two times scales.
If the leakage velocity function varies rapidly compared to the time evolution
 scale of unperturbed equation then it can be solved by perturbative method. In such case, the forcing functions become related self consistently with the 
 initial conditions. Plots of one lump solution in presence of such kind of leakage are shown. Since such kind of analytical treatment with nontrivial bottom boundary effects
 on superfluid  $^{4}He$  film is new as far as our knowledge goes it  may have its
useful applications and may pave new direction of research.

\section{Appendix : Exact lump wave solution of the forced KP-I equation (\ref{rrfKP}) for space time dependent forcing function $f(X,Y,T)$}
We can derive  one lump solution of our fKP equation (\ref{rrfKP}) following the method shown in \cite{Yong} if the function $f$, present in (\ref{rrfKP}) is written as $f = 8 (|\psi|^2)_X$ where the function $\psi$ satisfies the nonholonomic constraint (\ref{psi}). Then the fKP equation (\ref{rrfKP}) can be written as
\begin{align}
 &{} [ U_T + 6 U U_X + U_{XXX}]_X - 3 U_{YY} = -8 (|\psi|^2)_{XX}, \nonumber \\
 &{} i \psi_Y = \psi_{XX} + U \psi , \label{psi}
\end{align}
where subscripts in each function denote partial derivatives. Obviously the solution is relevant for a special physical condition when the function $\psi$ has dependence on spatial coordinates given by (\ref{psi}). 
Then the exact one lump wave solution $U$ of (\ref{psi}) along with the function $\psi$ can be written as
\begin{align}
&{} U = 2 [ ln (F) ]_{XX}, \ \  \psi =\frac{G}{F}, \ \  F = 1 + \xi_1^2 + \xi_2^2, \  \ G = G_R + i G_I, \nonumber \\
&{}  \ \xi_1 = a_1 X + a_2 Y + a_3 T + a_4, \ \ \  \xi_2 = a_5 X + a_6 Y + a_7 T + a_8,  \nonumber \\
&{} G_R = b_0 + b_1 \xi_1 + b_2 \xi_2 + b_3 \xi_1^2 + b_4 \xi_2^2, \nonumber \\
&{} G_I = c_0 + c_1 \xi_1 + c_2 \xi_2 + c_3 \xi_1^2 + c_4 \xi_2^2, \label{flump}
\end{align}
 with
 \begin{align}
  &{} a_3 = \frac{(a_2^2 - a_1^2)[3(a_1^4 + a_2^2 )^2 + 16 a_1^4]}{a_1(a_1^4 + a_2^2)^2}, \ a_5 = 0, \ a_6 = a_1^2, \ a_7 = \frac{2 a_1 a_2[3(a_1^4 + a_2^2 )^2 - 16 a_1^4]}{(a_1^4 + a_2^2)^2},
 \end{align}
where $a_1, a_2, a_4, a_8$ are arbitrary constant parameters.
Similarly we can get the other relations between the coefficients :
\begin{align}
  &{} b_0 = \frac{b_3(a_2^2 - 3a_1^4)}{(a_1^4 + a_2^2)}, \ b_1 = k c_1, \ b_2 = k c_2, \ b_4 = b_3,
 \end{align}
 
 \begin{align}
  &{} c_0 = -k b_0, c_1 = -\frac{4a_1^2a_2b_3}{(a_1^4 + a_2^2)}, \ c_2 = \frac{4a_1^4b_3}{(a_1^4 + a_2^2)}, \ c_3 = -k b_3, \ c_4 = c_3,
 \end{align}
 where $c_1, c_2, c_3, b_0, b_3, k$ are arbitrary parameters with $b_3^2 (1+k^2) = 1.$

\section{Acknowledgement}
The work has been carried out with partial financial support from the
Ministry of Science and Higher Education of the Russian Federation in the
framework of Increase Competitiveness Program of NUST $ «MISiS» ( K4-2018-061),$ implemented by a governmental decree dated 16 th of March 2013, N 211. Author acknowledges Prof. M. Lakshmanan, Bharathidasan University, India for the fruitful discussions. Author would also like to thank Ms. Aurna Kar for checking the grammar and literary aspects of the manuscript.


%

\end{document}